\documentclass[letterpaper, 10 pt, conference]{ieeeconf}  

\IEEEoverridecommandlockouts                        
\usepackage{amsmath}
\usepackage{amsfonts}
\usepackage{graphicx}
\usepackage{algorithm}
\usepackage[caption=false,font=footnotesize]{subfig} 
\usepackage[noend]{algpseudocode}
\usepackage{cite}
\usepackage{balance}
\usepackage{bm}
\usepackage{color}
\usepackage{float}
\usepackage{xfrac}
\usepackage{comment}
\usepackage[table,x11names,dvipsnames,svgnames]{xcolor}%
\usepackage{booktabs}
\usepackage{colortbl}
\usepackage{soul}
\usepackage{xurl}

\newtheorem{problem}{Problem}

\usepackage{multirow} 

\newcommand{\real}{\mathbb{R}}
\newcommand{\E}{\mathbb{E}}

\newcommand{\newcommenter}[2]{%
  \expandafter\newcommand\csname #1\endcsname[1]{%
    \fcolorbox{#2}{#2}{\color{white}\textsf{\textbf{#1}}}
    {\color{#2}##1}}%
  \expandafter\newcommand\csname at#1\endcsname{%
    \fcolorbox{#2}{#2}{\color{white}\textsf{\textbf{@#1}}}
    {\color{#2}}}%
  \expandafter\newcommand\csname #1hl\endcsname[2]{%
    \colorbox{#2}{\color{white}\textsf{\textbf{#1}}}\sethlcolor{Azure2}\hl{##2}~%
    \expandafter\ifx\csname commentarrow\endcsname\relax$\leftarrow$\else \commentarrow[#2]\fi~%
    {\color{#2}##1}}%
  \expandafter\newcommand\csname #1st\endcsname[2]{%
    \colorbox{#2}{\color{white}\textsf{\textbf{#1}}}\sout{##2}~%
    \expandafter\ifx\csname commentarrow\endcsname\relax$\leftarrow$\else \commentarrow[#2]\fi~%
    {\color{#2}##1}}%
}
\newcommenter{sba}{Purple}
\newcommenter{sw}{Red}
\newcommenter{zyc}{Brown}

\definecolor{DarkGray}{rgb}{0.19,0.31,0.31}

\title{\LARGE \bf
Multi-agent Robust and Optimal Policy Learning for Data Harvesting
}

\author{Shili Wu$^1$, Yancheng Zhu$^2$, Aniruddha Datta$^1$, and Sean B. Andersson$^{2,3}$
\\
$^1$Dept. of Electrical and Computer Engineering, Texas A\&M University, College Station, TX 77843, USA \\%
$^2$Dept. of Mechanical Engineering, $^3$Division of Systems Engineering,\\ 
Boston University, Boston, MA 02215, USA \\%
}

\begin{document}
\maketitle
\thispagestyle{empty}
\pagestyle{empty}

\begin{abstract}
We consider the problem of using multiple agents to harvest data from a collection of sensor nodes (targets) scattered across a two-dimensional environment. These targets transmit their data to the agents that move in the space above them, and our goal is for the agents to collect data from the targets as efficiently as possible while moving to their final destinations.  The agents are assumed to have a continuous control action, and we leverage reinforcement learning, specifically Proximal Policy Optimization (PPO) with Lagrangian Penalty (LP), to identify highly effective solutions. Additionally, we enhance the controller's robustness by incorporating regularization at each state to smooth the learned policy. We conduct a series of simulations to demonstrate our approach and validate its performance and robustness.
\end{abstract}

\section{INTRODUCTION}
Wireless Sensor Networks (WSNs) play a significant role in many applications, including environmental monitoring~\cite{HART:Reviews}, smart cities~\cite{Lei:WSNSmartCity} and surveillance~\cite{Zhijun:Motionplanningsurveillance}. WSNs are composed of a (relatively) large number of fixed sensor nodes deployed over a large geographical area. When it is impractical or infeasible to broadcast the sensor data out to users, a (relatively) small number of mobile agents, such as unmanned ground or aerial vehicles, can be used to move among the nodes and harvest their data, delivering it to a data sink. This approach can overcome energy, power, and computational limitations at the nodes~\cite{Moazzez:delay_data_harvesting}. At the same time, these agents have their own constraints, and as a result, it is important to determine efficient trajectories for the agents to perform their mission successfully.

When the problem takes place in one dimension, it is possible to derive a provably optimal controller~\cite{zhu:2022optimalcontroldataharvesting}. However, while there are applications where one dimension is sufficient~\cite{Zhou2017}, in most practical situations, the agents and sensor nodes operate in a higher dimensional space. While there has been significant effort in addressing optimal data harvesting in higher-dimensional spaces, it is a far more challenging problem. This is primarily due to the lack of a globally optimal parameterization of a controller that would allow for straightforward optimization~\cite{Lin:2015hn}. An alternative to the global parameterization approach is to formulate the problem as a Combinatorial Optimization (CO) problem. In this approach, the agent performs rollouts using a model and applies a search technique to identify the optimal action \cite{darvariu2024graph}. However, the high dimensionality of the data harvesting problem makes it computationally infeasible for the agent to explore all possible paths, as the time complexity of such an algorithm is $O(b^H)$, where $b$ represents the size of the action space and $H$ is the planning horizon. To overcome this challenge and reduce the complexity, a more feasible strategy is to use random rollouts combined with a reinforcement learning approach \cite{bayerlein2021multi, swu2023TimeOptimal}. In our previous work, we developed an RL-based controller that directly optimized the binary reward function~\cite{swu2023TimeOptimal}. This approach allowed us to avoid the need for reward engineering and reliance on intuitive approaches to formulate the reward function to produce a ``desired result'' that is not encoded in the rewards. It was limited, however, to a single agent with a finite number of available control actions and ignored issues of measurement noise, modeling error, and disturbances. In practice, the use of multiple agents can be beneficial, especially when the WSN is distributed over a large area and the control actions are better modeled as bounded and continuous. In this paper, we build upon our previous effort, allowing multiple agents to do data harvesting simultaneously and move away from a discrete representation of the available controls into a continuous action space.


When considering data harvesting in practical scenarios, it is important to recognize that the system is subject to noise and disturbances, both in the measurements and in the control, arising from environmental factors and modeling errors. Particularly when applying learning-based methods, these disturbances can lead to undesirable behavior, such as policy vibrations, that lead to poor performance. One can address this in the learning process by, for example, modifying the critic in an actor-critic formulation to account for the noise and ensuring disturbances are modeled in the sampling process during learning~\cite{wan2020robust}. However, these approaches typically assume a Gaussian distribution on the noise, an assumption that may not hold in practice. We take a different approach based on network smoothing. Smoothing has been shown to be an effective method for finding a robust estimator in supervised learning~\cite{DBLP:journals/corr/abs-1808-09540}, as well as for finding a robust controller in trajectory tracking tasks \cite{pmlr-v202-song23b}.


Our goal in this paper is to find a control policy over a continuous action space for a collection of agents so as to harvest all the data from a set of target sensor nodes in a two-dimensional environment in minimal time and in a way that is robust against disturbances. The primary contributions of our work are: 
\begin{itemize}
    \item A deep reinforcement solution to the time-optimal, two-dimensional, multi-agent data harvesting problem.
    \item An exploration of the trade-off between different terminal penalty assignments that shows the advantage of using a Lagrangian Penalty in terms of finding a higher-performing solution.
    \item A learned policy that exhibits robustness to disturbances through the use of regularization for policy smoothing.
\end{itemize}
Through simulations, we demonstrate improvement over both our prior work and a classical shortest path finding algorithm as a baseline. We also demonstrate the robustness of the scheme with respect to disturbances in the measured state. Note that while we focus on the data harvesting problem, our approach can also be applied to similar multi-agent control problems like pickup and delivery or persistent monitoring.


\section{Problem Formulation \label{sec:prob}}
\subsection{System Design}
\begin{figure}[htbp!]
	\centering
	\includegraphics[width=0.45\textwidth]{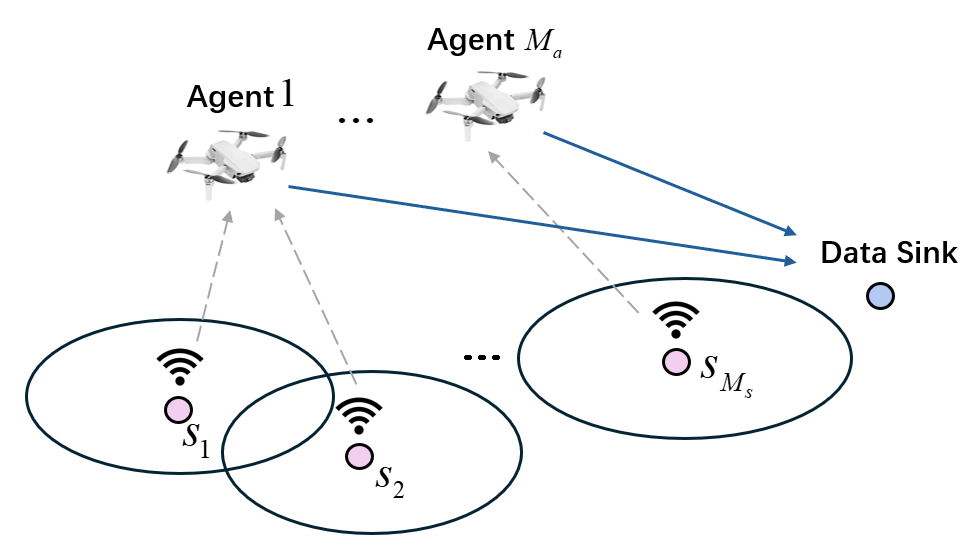}
	\caption{Illustration of the data harvesting problem. Multiple agents, flying at constant heights, harvest data from sensor nodes to bring the information back to data sinks for use. The circles represent the effective communication range.}
	\label{fig:2d-working-space-demonstration}
\end{figure}

As illustrated in Fig.~\ref{fig:2d-working-space-demonstration}, we consider a collection of $M_s$ sensor nodes (also referred to as \textit{targets}) located at positions $s_i\in \mathbb{R}^2$, $i=1,2,\dots,M_s$. The initial data volume to be harvested from each target $i$ is given by $D_i \in \mathbb{R}_{\geq 0}$ where $\mathbb{R}_{\geq 0}$ denotes the set of non-negative real numbers. The joint initial data volume collection is denoted as $\mathbf{D} \in \real^{M_s}_{\geq 0}$. Harvesting is performed by $M_a$ agents flying at a constant height $h$ over the workspace. Agent $j$ ($j=1,\dots,M_a$) starts at location $p_{start}^j \in \mathbb{R}^2$, and must end at $p_{final}^j \in \mathbb{R}^2$. For example, the final location may be at a data sink so that the agent can offload the acquired data. The joint position of the agents is given by $\mathbf{p} \in \mathbb{R}^{2M_a}$.

We model the data transmission rate $\psi_{ij}$ between target $i$ and agent $j$ by combining the Shannon-Hartley Theorem~\cite{Shannon:1949} with the Friis Transmission Formula~\cite{Friis:1946}, yielding
\begin{equation}
    \psi_{ij} (p_j,s_i)={B} \log_2\left(1 + \frac{K}{ \parallel p_j - s_i \parallel_2^2}\right),
    \label{eq:Phi_Model}
\end{equation}
where $B$ is the bandwidth of the channel, and $K$ is a constant proportional to the output power of the transmitting antenna, the wavelength of the radio frequency, and the gains of the antennas. See \cite{zhu:2022optimalcontroldataharvesting} for details. The amount of data harvested at time step $n$ is denoted as $d_i(n)$, and the combined joint vector is denoted as $\mathbf{d}(n) \in \mathbb{R}^{M_s}_{\geq 0}$. The goal is to minimize the time it takes for the agents to move from $\mathbf{p}_{start}$ to $\mathbf{p}_{final}$ while harvesting all the data from all the targets.

We assume a discrete-time formulation over $N$ time steps where $N$ is not fixed \textit{a priori} but is the outcome of the problem solution. At each time step $n$, the agents move according to the model
\begin{align} \label{eq:posUpdate}
    \mathbf{p}(n+1) = \mathbf{p}(n)+ \mathbf{a}(n),
\end{align}
where $\mathbf{a}(n )\in \mathcal{U} \subset \real^{2M_a}$ is the joint action for all the agents and $\mathcal{U}$ is the admissible set of controls. In this work, we represent the action for each agent in polar form $(\rho_i,\alpha_i)$ and define the admissible set as $\rho_i \in [0, \rho_{\max}]$, $\alpha_i \in [0,2\pi)$. 


The change in the data received from the $i^{th}$ target is modeled as
\begin{align} 
    &d_i\left( n+1 \right) = d_i \left(n\right) + \Delta d_i(\mathbf{p}(n), \mathbf{a}(n)), \label{eq:dataUpdate}
\end{align}
where $\Delta d_i$ is determined by accumulating the data according to the transmission rates as the agent moves from $\mathbf{p}(n)$ to $\mathbf{p}(n+1)$ according to the action taken $\mathbf{a}(n)$.
We define the system state $x_n$ to be the concatenation of the agent's positions and the data acquired, $x_n = \begin{bmatrix} \mathbf{p}_n^T & \mathbf{d}_n^T\end{bmatrix}$. The overall state space is denoted as $\mathcal{X}$. The combination of \eqref{eq:posUpdate} and \eqref{eq:dataUpdate} defines the state update $\mathcal{P}$.

\subsection{Optimization Problem with Lagrangian Penalty}
The overall goal of the task is to minimize the duration required to reach the goal state, defined as 
\begin{align}
    g = \begin{bmatrix}
        \mathbf{p}_{final}^T & \mathbf{D}^T
    \end{bmatrix}^T.
\end{align}


We reframe the problem as maximizing a cumulative return $J(\tau)$ of a given trajectory $\tau$ by assigning a reward of $r(x) = -1$ at each time step until the terminal state is reached. With this, the problem is a Markov Decision Process (MDP) as defined by the elements $(\mathcal{X}, \mathcal{A}, \mathcal{P}, r, \gamma)$.

As is typical in RL, the system is not guaranteed to meet the final constraints during training. To effectively search for an optimal policy, we apply a Lagrangian penalty method, a term borrowed from optimization theory due to the structure of the new reward function to the terminal state of the trajectory, leading to the following optimization problem.

\begin{problem}\label{prob:Lagrangian}
\begin{equation*}
    \begin{array}{cl}
        \max_{\pi} & J(\tau) = \sum_{n=0}^{N-1} \gamma^nr(x_n) - \lambda^Tf(x_N, g)\\ \\
        {\text{subj. to }} & x_0 = \begin{bmatrix} \mathbf{p}_{start}^T & \mathbf{0}^T_m \end{bmatrix}^T, \\
        & x_{n+1} = \begin{bmatrix} \eqref{eq:posUpdate} & \eqref{eq:dataUpdate} \end{bmatrix}^T, \\
        & a_n = \pi(x_n) \in \mathcal{U},
    \end{array}
\end{equation*}
\end{problem}
\noindent where $\pi(\cdot)$ is the policy mapping from the state to action, $\lambda$ is called the Lagrange multiplier vector and $f: \mathbb{R}^{2M_a + M_s} \mapsto \mathbb{R}^{M_\lambda}$ (where $M_\lambda$ is the number of terminal constraints) encodes the penalty for not completing the task. In our setting, we have two terminal constraints: one on reaching the terminal location (where we take the constraint penalty to be the maximum time among the agents to reach their terminal location moving at maximum speed), and one on collecting all the data (where we take the constraint penalty to be the 1-norm on the data remaining at targets at the end of the trajectory). Choosing the Lagrange multipliers sufficiently large ensures that unsuccessful trajectories receive lower rewards than successful ones and that the rewards increase as the trajectory gets closer to satisfying the constraints.
Moreover, the two problems (original and with Lagrangian penalty) have the same optimal solution, and thus, solving the problem with penalties can give us the solution to the original problem. In the remainder of this work, we use a policy gradient method (described in Sec. ~\ref{sec:methodology}) to find a solution to the Lagrangian formulation.


When applying the optimal policy, the system may receive incorrect state information $\hat{x}$ due to measurement noise or errors in the data transmission. 
Consequently, the controller may output a sub-optimal action $\pi(a|\hat{x})$ based on this disturbance rather than the optimal action $\pi(a|x)$ determined by the true state. As a result, the performance of the controller might be significantly affected. We, therefore, seek a solution that is robust against such disturbances.

\section{methodology}\label{sec:methodology}

We take a centralized approach to our problem (leaving the problem of a decentralized solution as future work) and seek a stochastic policy. That is, recalling that the control action for agent $j$ is in the polar form $(\rho_j, \alpha_j)$, the values at each step are sampled from the normal distributions
\begin{align}
    \rho_j \sim \mathcal{N}\left(\mu^{\rho}_j(x),\sigma^{\rho}_j(x)\right), \quad
    \alpha_j \sim \mathcal{N}\left(\mu^{\alpha}_j(x),\sigma^{\alpha}_j(x)\right).
\end{align}
Finding an optimal policy means finding the best (state-dependent) mean and variance parameters for these distributions. Using a stochastic policy during training enables an effective exploration of the state and action space to drive learning. At test time, however, we use a deterministic policy by simply selecting the means of the distributions.

\begin{figure}[htbp!]
	\centering
	\includegraphics[width=0.48\textwidth]{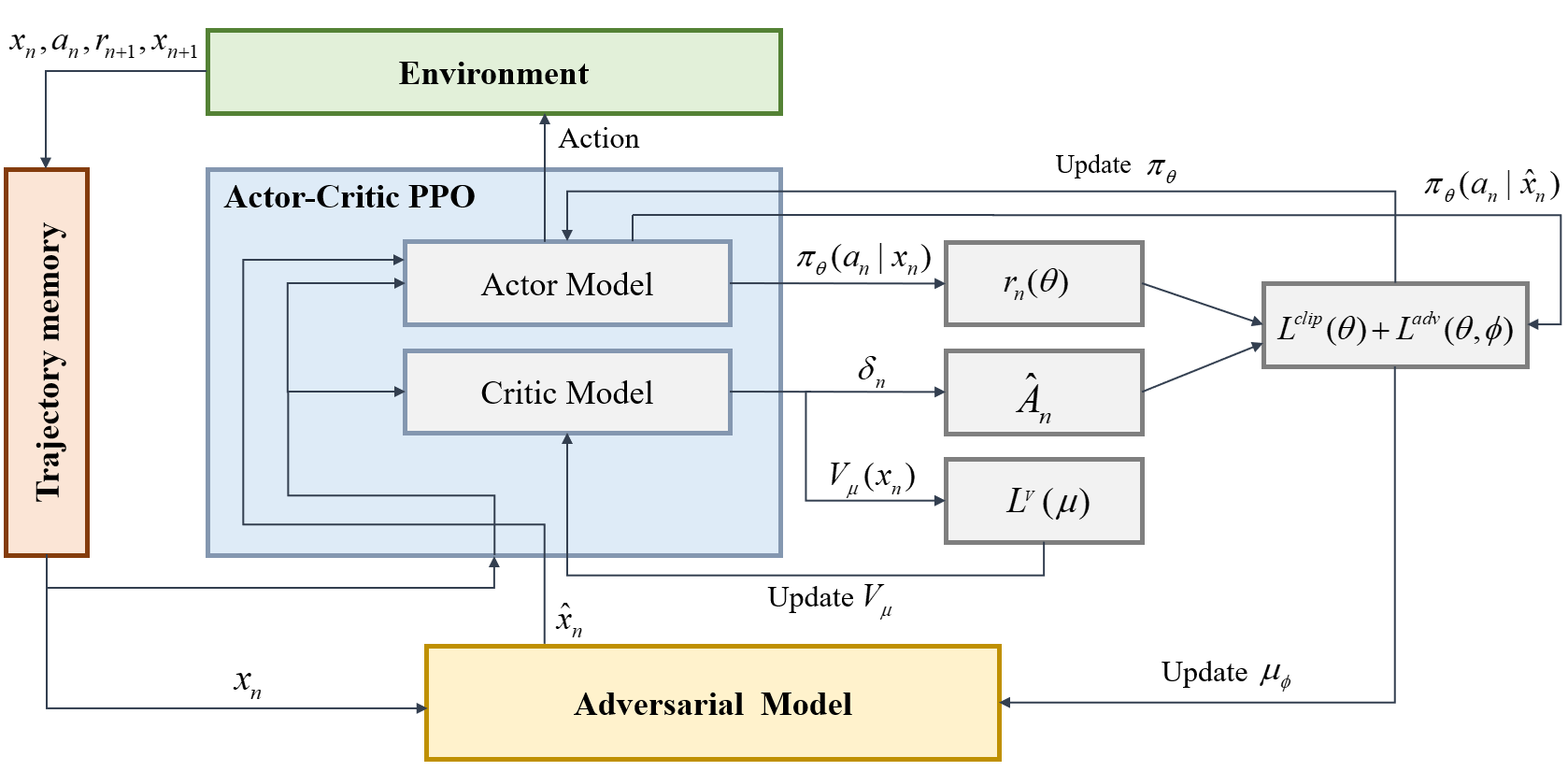}
	\caption{Algorithm Framework}
	\label{fig:algorithm structure}
\end{figure}

To handle the continuous action and state space, we turn to Proximal Policy Optimization (PPO)~\cite{schulman2017proximal} to learn our controllers; this is described in Sec.~\ref{sec:PPO}. To make these controllers robust to uncertainty and noise, we introduce regularization by adding a penalty for divergent actions taken in response to erroneous states; this is described in Sec.~\ref{sec:regularlization}. Our overall approach is diagrammed in Fig.~\ref{fig:algorithm structure}. 

\subsection{Proximal Policy Optimization \label{sec:PPO}}


While a variety of gradient-based methods can be used to optimize our policy, we select PPO as it has been shown in the literature to have good performance, is easy to implement, and is straightforward to tune. PPO follows the general principle of policy gradient and uses an actor-critic structure. The critic network provides an estimate of the value function $V$, which can be interpreted as the approximate completion time at each state under the current policy $\pi$. The actor-network then optimizes its policy to maximize the expected return by adjusting its actions to increase the value at each state. The details on PPO can be found in a variety of sources, including \cite{schulman2017proximal}.
\subsection{Policy Network Smoothing \label{sec:regularlization}}

To develop a robust solution, we introduce a regularizer to the policy network that aims to smooth the output of the policy network, encouraging the generation of similar actions when confronted with perturbed states \cite{shen2020deep}. The regularization approach, known as $SR^2L$ \cite{shen2020deep}, begins by defining a perturbation set $\mathcal{B}_d (x, \epsilon) = \left\{\hat{x}: d(\hat{x}, x) \leq \epsilon\right\}$, where $d$ represents an $l_p$ metric defining a notion of distance on the state space. It then uses an adversarial neural network $\mu_\phi(x): \mathcal{X} \rightarrow \mathcal{X}$ parameterized by $\phi$ to identify a perturbed state $\hat{x} \in \mathcal{B}_d (x,\epsilon)$ that maximizes the distance $\mathcal{D}$ between the actions of the actor $\pi_\theta$ before and after the perturbation. Such identification is achieved by solving the following maximization problem
\begin{align}
    L_\phi = \max_{\mu_\phi(x) \in \mathcal{B}_d (x, \epsilon)} \mathcal{D} \left(\pi_\theta(\cdot|x), \pi_\theta(\cdot|\mu_\phi(x))\right),
\end{align}
where $L_\phi$ defines the loss function for training $\mu_\phi$. After the adversarial $\mu_\phi$ is learned at each training step, we add a regularization term as an auxiliary loss for the actor-network:
\begin{align}
     R_x^\pi = \E_{x \sim \rho^\pi} \mathcal{D} \left(\pi_\theta(\cdot|x), \pi_\theta(\cdot|\mu_\phi(x)\right).
     \label{eq: regularizer}
\end{align}




When the actor minimizes with respect to this auxiliary loss, it encourages a similar action choice in nearby states, making the policy ``smoother''. In Sec.~\ref{sec:sims}, we refer to the original policy as ``non-smooth'' and the one trained using the adversarial network as the ``smoothed'' policy.


\section{simulation}\label{sec:sims}
\subsection{Simulation Set-up}
To demonstrate our approach, we consider two simulation settings. The first has four targets placed at $(3,1)$, $(7, 1)$, $(7,5)$ and  $(7,7)$, and three agents, all starting at $(0,1)$ and ending at $(7,9)$. This somewhat simple configuration has the advantage of having an intuitive optimal (or at least high-performing) solution. The transmission parameters were set to $K=0.7$ for all targets and $B = [0.5,1.0,1.5,2.0]$ for each target respectively. All agents were kept at a height of $h = 0.5$. The initial data volumes were set to $\mathbf{D} = \begin{bmatrix} 5 & 6 & 3 & 3 \end{bmatrix}^T$.

The second configuration involves five targets, placed at $(3,1)$, $(6,7)$, $(8,2)$, $(1,6)$ and $(3,9)$ with initial data volumes of $\mathbf{D} = \begin{bmatrix} 5 & 6 & 3 & 4 & 4\end{bmatrix}^T$. There were once again three agents, now all starting at the origin but having final locations of $(9,6)$, $(5,5)$, and $(7,8)$. The communication parameters were set to $K=0.7$ and $B = [0.5,1.0,1.5,2.0,2.5]$.

Before each learning episode, the agents were given a rough estimate of the possible completion time and were allowed to execute a maximum of $N_{max}$ steps. To compare the performance of different approaches, if the agents were unable to complete the task during evaluation, we calculated the expected completion time as the sum of the execution time, plus the additional time the agents needed to spend in the terminal state of the evaluation to complete data collection and the time needed for all agents to move from their terminal positions to their desired target locations assuming they moved directly and at maximum speed.


In the following sections, we first show the advantage of using a continuous action space over a discrete action space. We then explore the advantage of using the Lagrangian penalty over other terminal penalization methods. Then, we demonstrate how a smooth regularizer helps to improve the robustness of the approach.

\begin{figure*}[!t]
    \centering
    \subfloat[Trajectory Learned by A*\label{fig:Case-B-2d-simulation}]{
        \includegraphics[width=0.3\textwidth]{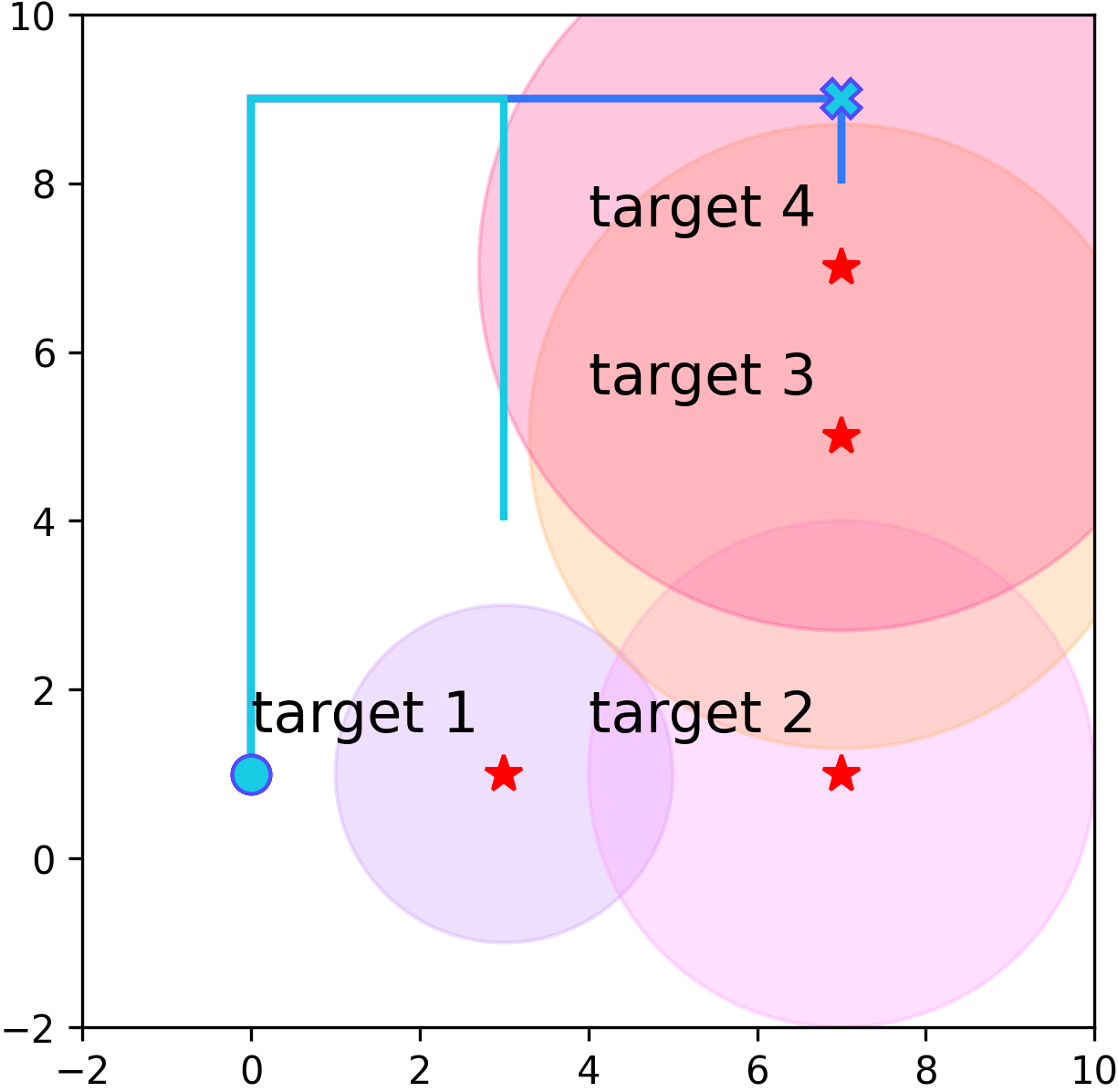}
    }\hfill
    \subfloat[Trajectory Learned by DDQN\label{fig:Trajectory-DDQN}]{
        \includegraphics[width=0.3\textwidth]{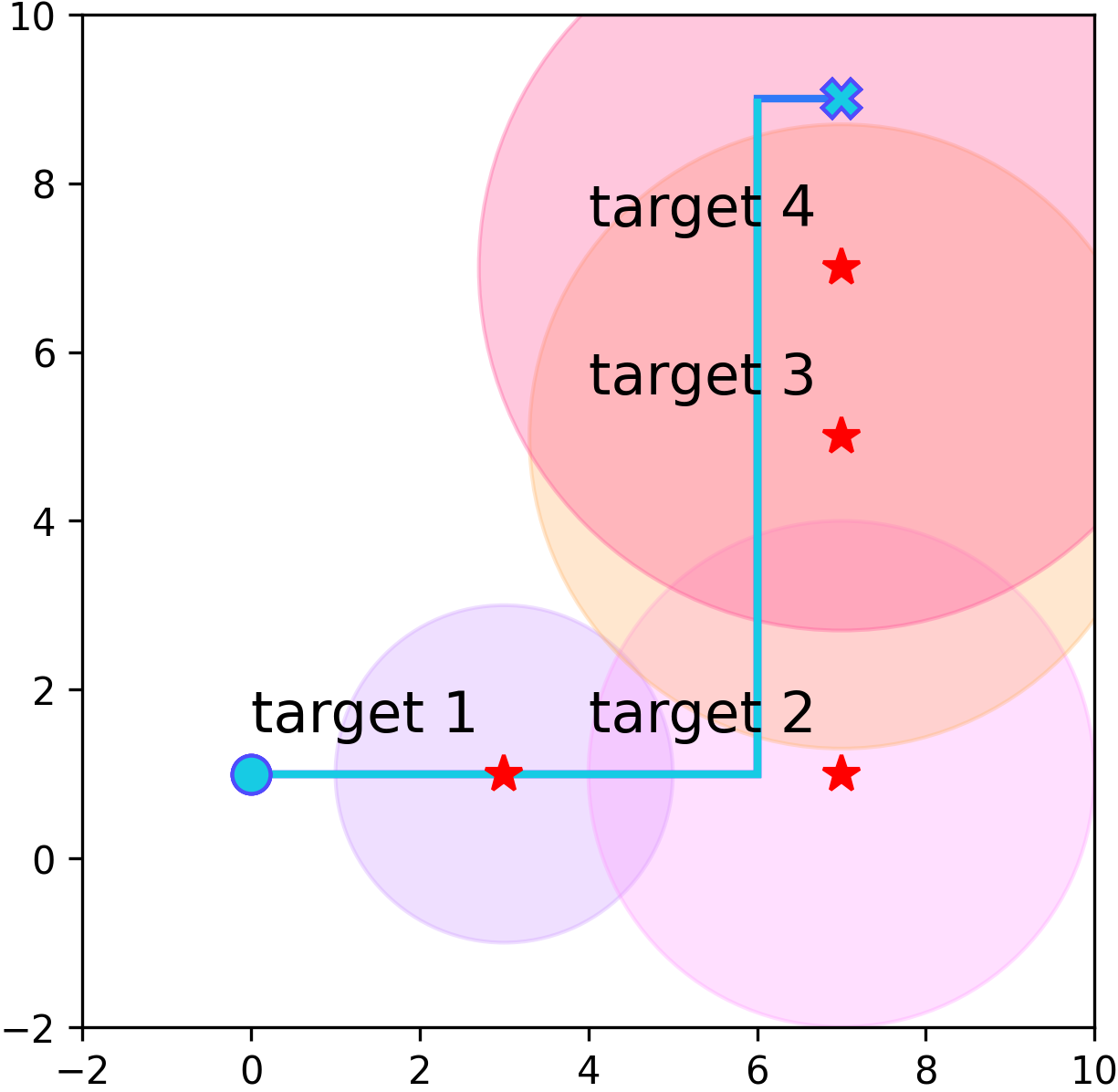}
    }\hfill
    \subfloat[Trajectory Learned by PPO\label{fig:Trajectory-PPO}]{
        \includegraphics[width=0.3\textwidth]{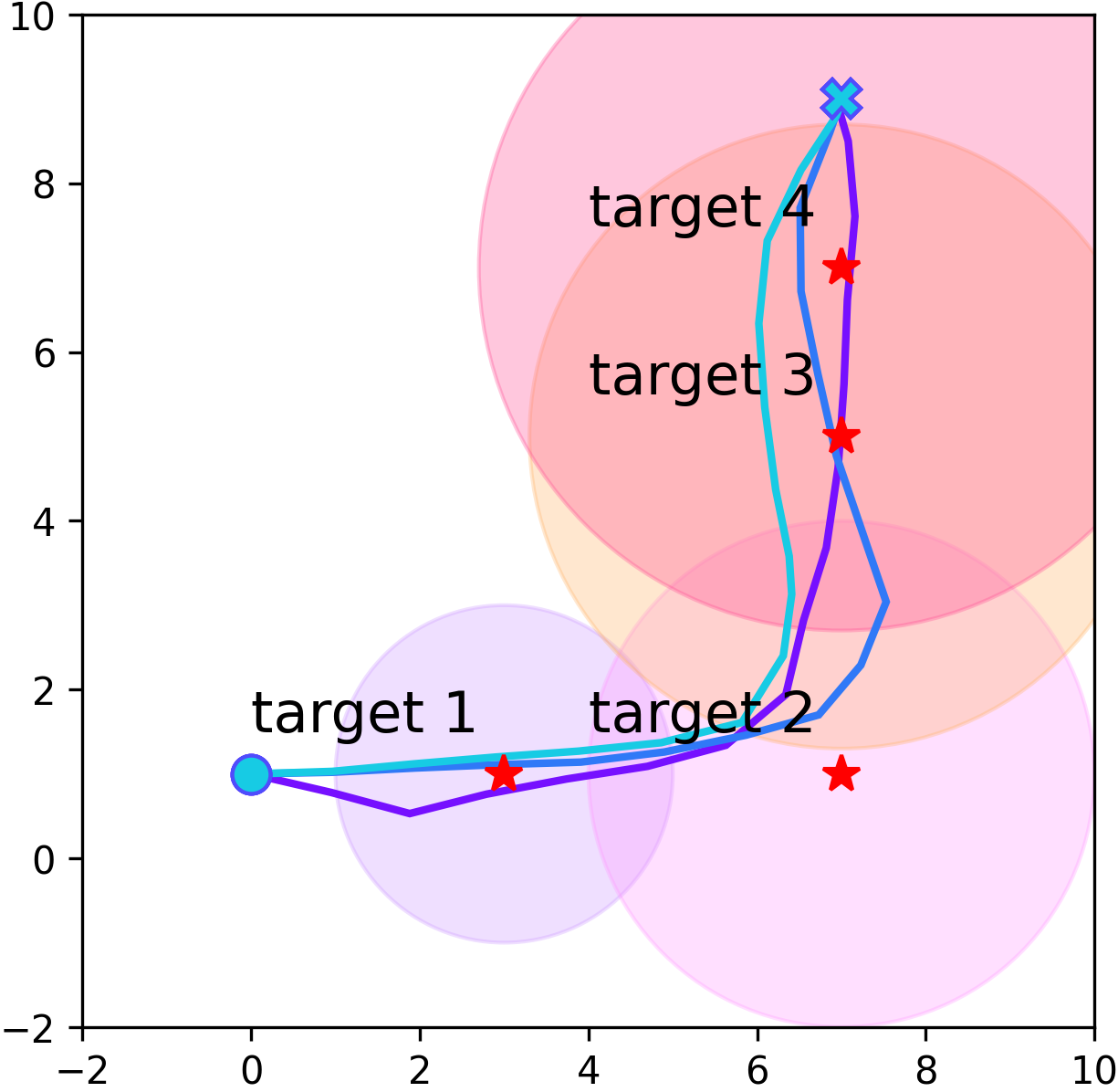}
    }
    \caption{Optimal trajectories in the four-target configuration. Agents all begin at the small blue circle. The terminal location is indicated by a blue \texttt{x}. Target positions are indicated by red stars, and the shaded circles represent the effective communication range (rate $\geq 0.01$). Trajectories obtained by (a) A*, (b) DDQN, and (c) PPO.}
    \label{fig:Optimal-Trajectory}
\end{figure*}

\begin{figure*}[!t]
    \centering
    \subfloat[Trajectory Learned by A*\label{fig:Case-B-2d-simulation-5t}]{
        \includegraphics[width=0.3\textwidth]{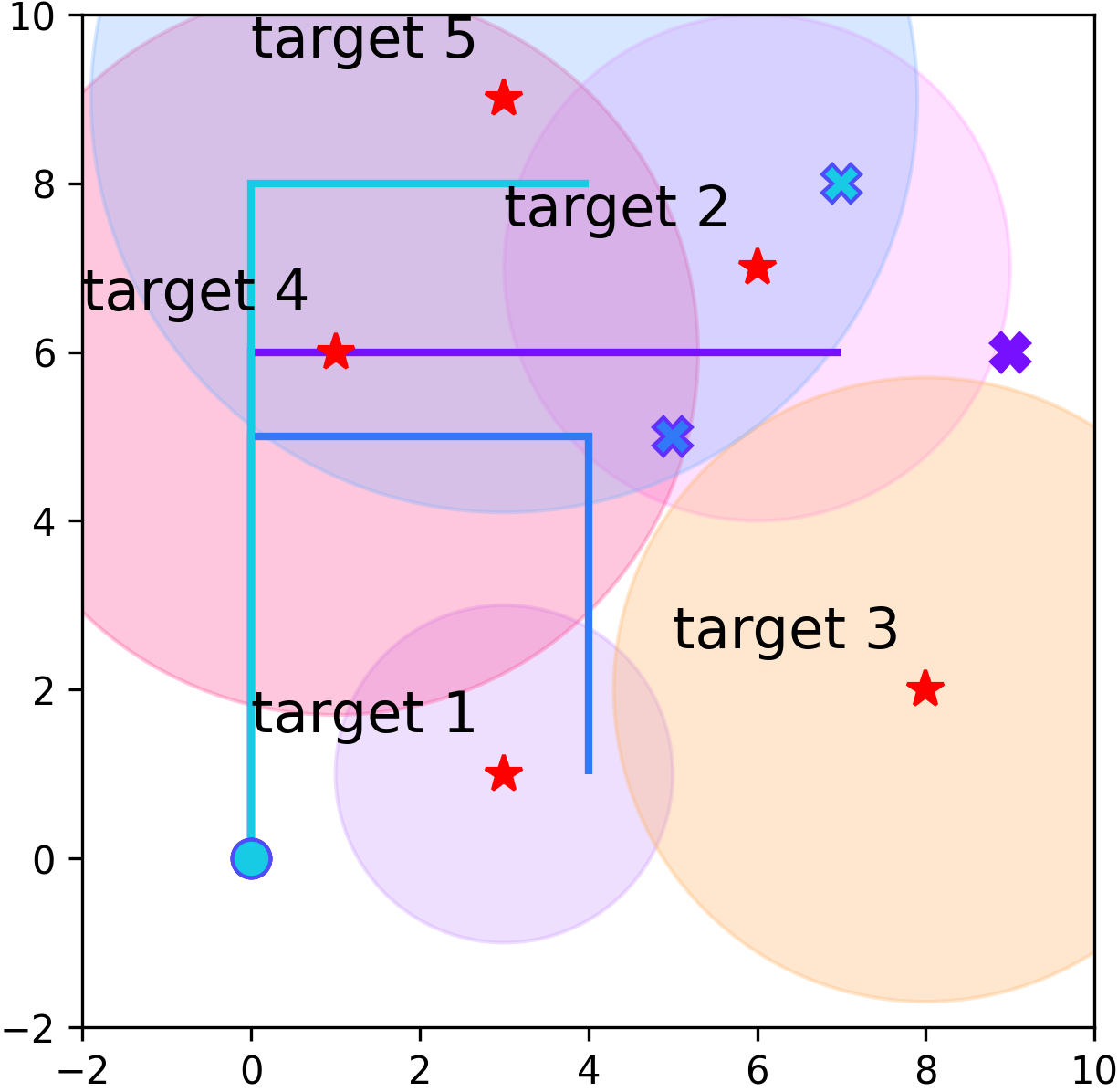}
    }\hfill
    \subfloat[Trajectory Learned by DDQN\label{fig:Trajectory-DDQN-5t}]{
        \includegraphics[width=0.3\textwidth]{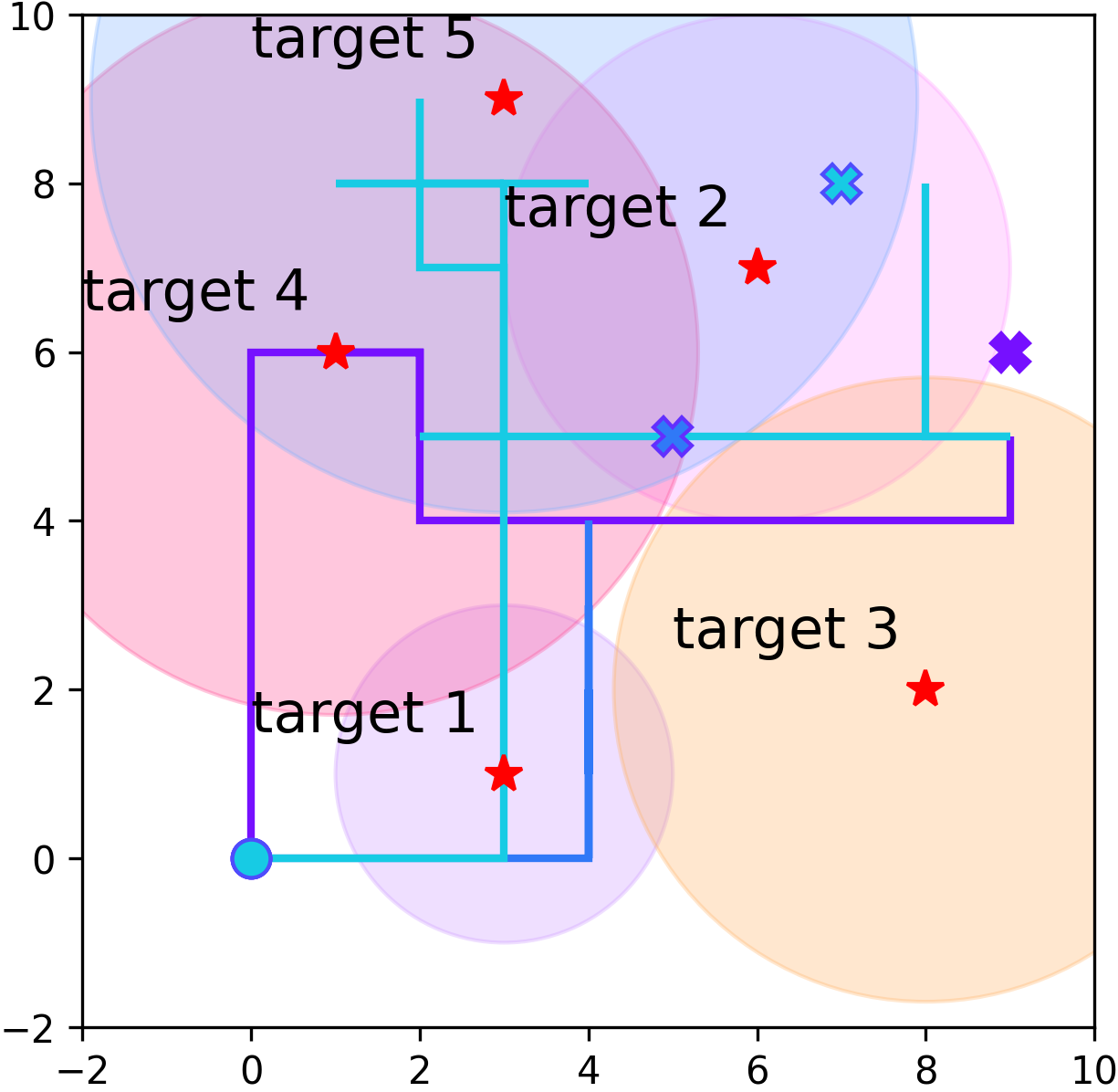}
    }\hfill
    \subfloat[Trajectory Learned by PPO\label{fig:Trajectory-PPO-5t}]{
        \includegraphics[width=0.3\textwidth]{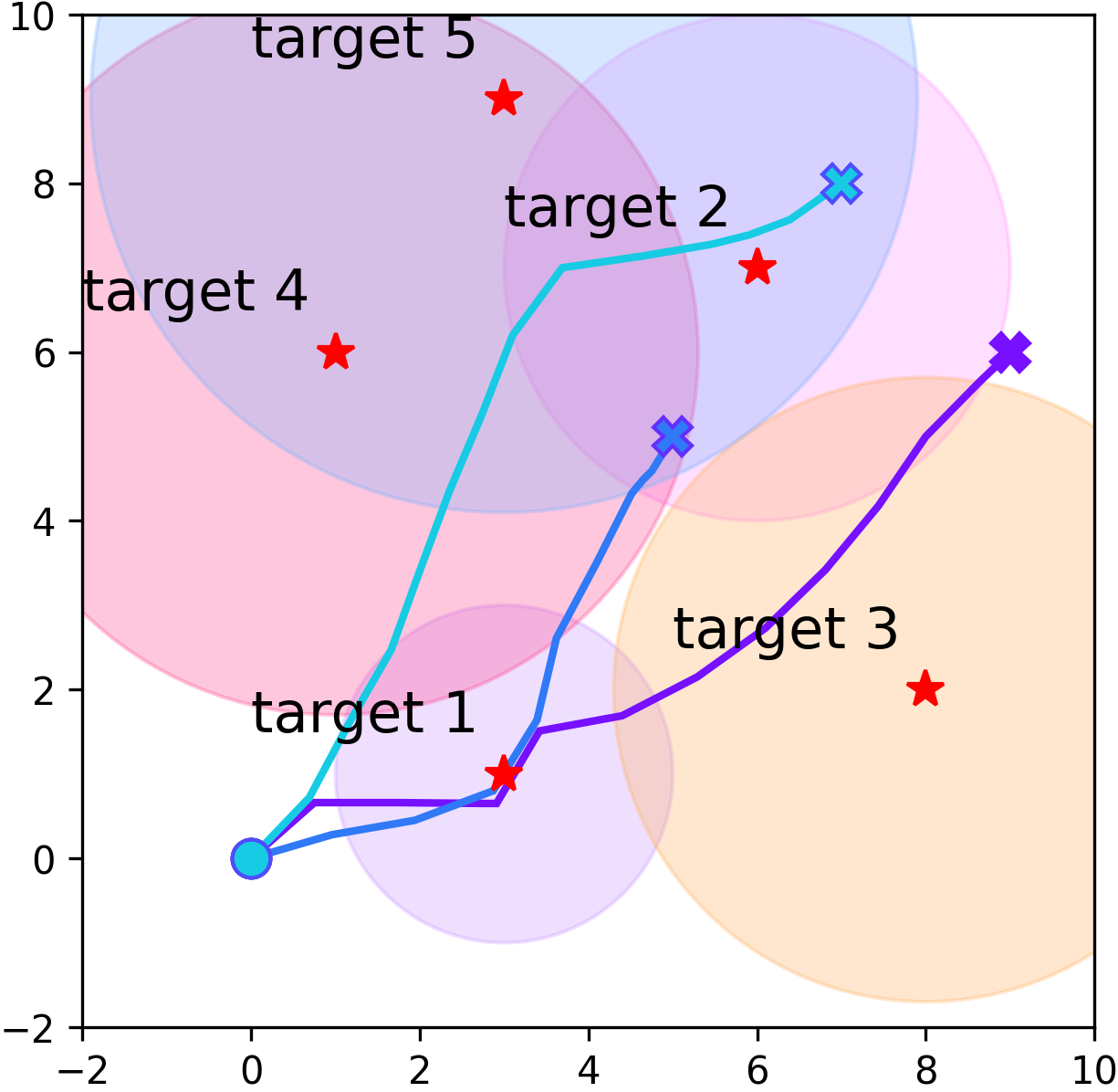}
    }
    \caption{Optimal trajectories in the five-target configuration. Agents all begin at the blue dot. Three terminal locations are shown using \texttt{x}'s. Target positions are indicated by red stars, and the shaded circles represent the effective communication range (rate $\geq 0.01$). Trajectories obtained by (a) A*, (b) DDQN, and (c) PPO.}
    \label{fig:Optimal-Trajectory-2}
\end{figure*}

\begin{figure}[!t]
    \centering
    \subfloat[Completion Time of Each Target (Configuration 1)\label{fig:task-completion-1}]{
        \includegraphics[width=0.48\textwidth]{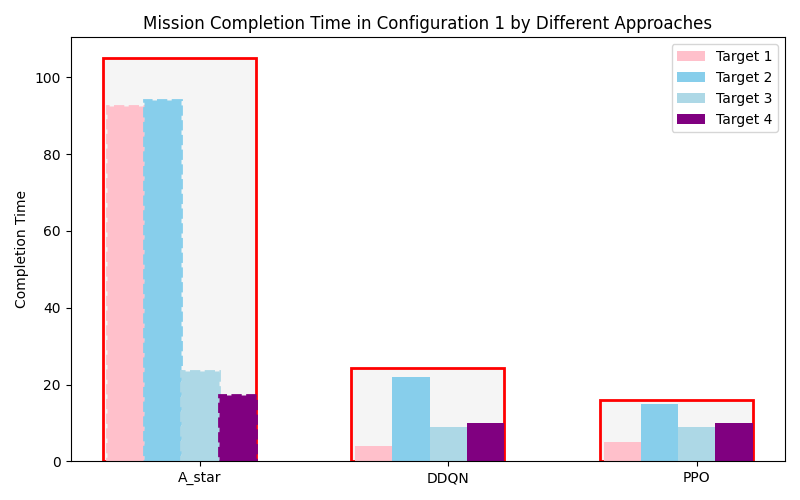}
    }\hfill
    \subfloat[Completion Time of Each Target (Configuration 2)\label{fig:task-completion-2}]{
        \includegraphics[width=0.48\textwidth]{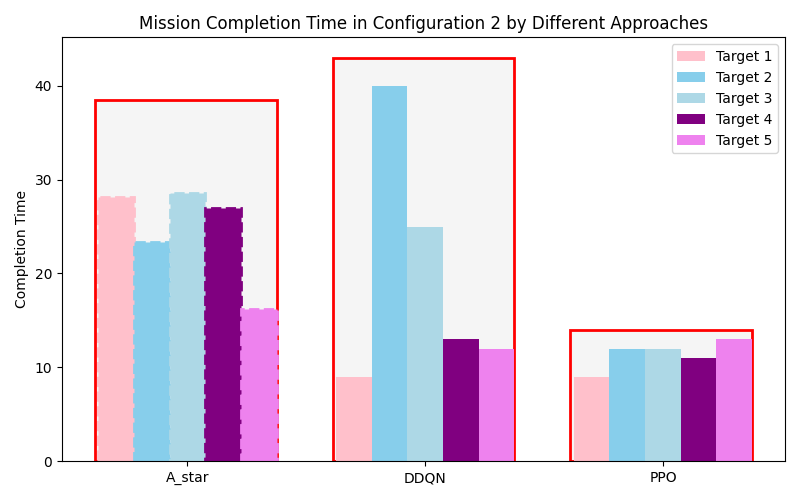}
    }
    \caption{Completion time of data collection from each target under two different scenarios. The red box indicates the final expected task completion time, including travel to the final destination. A dashed outline indicates cases where data collection was not completed during policy execution.}
    \label{fig:Expecting-Completion-Time}
\end{figure}

\subsection{Optimal Trajectory}

To illustrate the performance benefits of using Proximal Policy Optimization (PPO) with a continuous action space, we compare it to Double Deep Q-Network (DDQN), as utilized in our previous work \cite{swu2023TimeOptimal}. This approach discretizes the action space, allowing the problem to be modeled as a graph where the nodes represent states and the edges form between pairs of nodes that can be reached by taking an action from the available action set. This graph-based formulation enables the use of the A* algorithm, a well-known method for finding the shortest path in a graph. A* employs a heuristic to guide the search process, making it more efficient than a standard breadth-first search. The heuristic implemented in this work at a given state is the lower bound of the actual completion time, which is the maximum between the time for the agents to move directly to the destination at maximum speed and the time to complete data, assuming a maximal transmitting rate. 

In the discrete action setting, the size of the joint action space in a multi-agent system grows exponentially with the number of actions available to each agent, and the size of the graph expands exponentially with both the joint action space and the planning horizon $H$. To manage this complexity, we limit the discrete action space for each agent to $\mathcal{A} = \{\textrm{north, east, south, west, hover\}}$ and opt for larger action steps to keep the control horizon short, thus making the computational complexity of both DDQN and A* manageable. However, this choice introduces a tradeoff, as larger action steps may lead to suboptimal solutions. In contrast, PPO, with its continuous action space, mitigates this challenge by allowing finer control over actions.

Typical runs of each approach in the two configurations are shown below with the trajectories in Figs.~\ref{fig:Optimal-Trajectory},~\ref{fig:Optimal-Trajectory-2} and the time to harvest the data for each target in Fig.~\ref{fig:Expecting-Completion-Time}. Consider first the four target settings (Figs.~\ref{fig:Optimal-Trajectory} and \ref{fig:task-completion-1}). Due to the size of the graph, A* was unable to find a solution within the search time allowed, ending with a best (in terms of expected completion time as defined above) solution of $105.39$ units of time. Under DDQN, all agents followed the same trajectory, completing the mission in 25 units. Under the PPO-based policy, the agents took slightly different paths and completed the mission in 18.02 units. Both sets of trajectories closely align with the intuitive solution where the agents move along a path of high transmission rate. Given the similarity in trajectories, the improved performance is almost certainly due to the additional flexibility of the continuous action space enabled in part by using PPO to solve the problem.


The performance on the five target configuration is shown in Figs.~\ref{fig:Optimal-Trajectory-2} and \ref{fig:task-completion-2}. In this configuration, there is a significant difference in the trajectories and in the time to complete the mission. In the particular run shown, the A* and DDQN approach failed to complete the mission within $40$ time units, with DDQN performing slightly worse (in terms of expected completion time) than A*. The final policy found by PPO took $14.02$ units to complete the mission.


\subsection{Effects of Cost Structure}

\begin{figure}[h]
\vspace{2pt}
	\centering
	\includegraphics[width=0.4\textwidth]{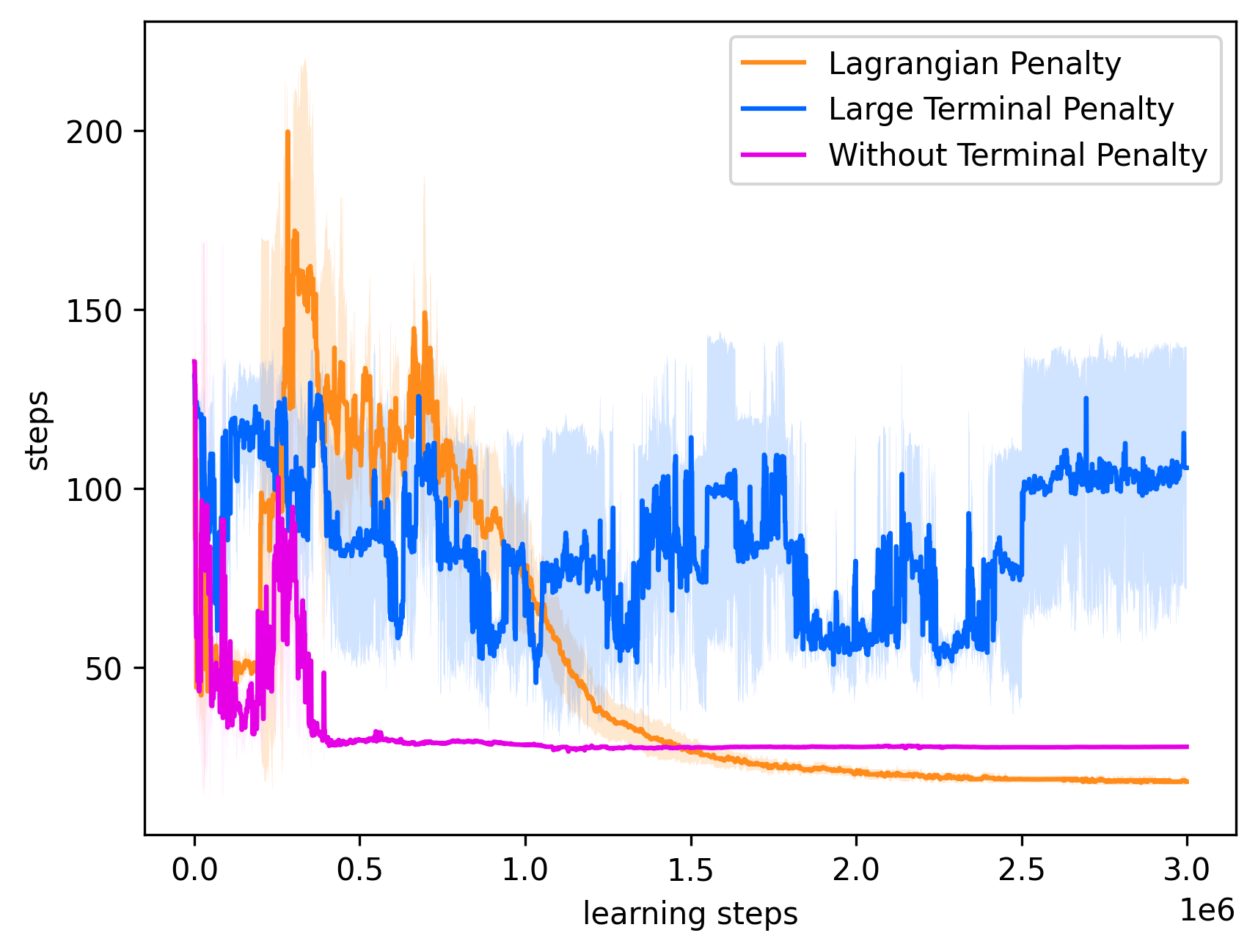}
	\caption{Comparing performance using (orange) Lagrangian Penalty, (blue) large, and (purple) no terminal penalty. Each learning step corresponds to one gradient update in PPO.}
        \label{fig:effects of Lagrangian}
 \end{figure}
 
To show the advantage of the Lagrangian penalty, we compared it against two other approaches, one that assigns a large penalty to unsuccessful trajectories
and one that ignores the terminal constraints entirely. 
For a fair comparison, all were solved using PPO. Since the cost functions are different, we use the number of steps to complete the problem (with a time of $N_{max}$ if the agents fail to extract all the data and move to their terminal locations) as the comparison metric. In each case, we initialized the learning scenario with $10$ random seeds and $3$ different sets of hyperparameters to create different training evolutions. 
The completion time as a function of the learning iteration is shown in  Fig.~\ref{fig:effects of Lagrangian} where the shaded region indicates the variance of performance over 30 trials. The final performance is listed in Table~\ref{tab:Diff Reward Design}. Using a large terminal penalty (blue solid line) shows poor convergence and performance relative to the other methods. While the lack of a terminal condition (solid purple line) led to the fastest convergence, using the Lagrangian penalty (solid orange line) led to the best final performance.


\begin{table}[htbp!]
    \centering
    \setlength{\tabcolsep}{13.0pt} 
    \renewcommand{\arraystretch}{1.1} 
    \scalebox{1.35}{
    \begin{tabular}{|c|c|}
        \rowcolor[HTML]{b1457c}  \textcolor{white}{Algorithm} & \textcolor{white}{Final Steps}  \\ \hline
        Lagrangian Penalty & 18.08 $\pm$ 0.845\\ \hline
        Large Terminal Penalty & 105.79 $\pm$ 33.8 \\ \hline
        Without Penalty & 27.84 $\pm$ 0.665\\ \hline
    \end{tabular}}
    \caption{Comparison of final performance from different reward schemes used on PPO framework} \label{tab:Diff Reward Design}
    \label{table:timings}
\end{table}

\subsection{Robust Policy}
To demonstrate the robustness of incorporating the regularizer, we subjected the agents to a bounded disturbance of size $\epsilon$ affecting both the communication rate from each sensor at each time step and the location of each agent as received by the centralized controller. During training, the disturbance set in the regularization was defined as $| \hat{x} - x |_{\infty} \leq 0.05$. We trained 5 models with different training parameters and evaluated their performance under two types of noise: ``random'' and ``adversarial''. During evaluation, agents observed perturbed states at every step and executed actions based on these perturbed states. The random perturbations were generated from a uniform distribution, while the adversarial noise was produced by the trained adversarial models coupled with $SR^2L$. We evaluated the models at noise levels $\epsilon = [0.025, 0.05]$, with each model tested 100 times to ensure accuracy. The performance results under varying types and magnitudes of perturbations are presented in Table~\ref{table:timings}. As observed, the adversarial model generally introduced stronger noise. Although the smooth policy achieved slightly worse results when no noise was presented, it exhibited slightly better performance when the noise was presented during the operation.

\begin{table}[htbp!]
    \centering
    \scalebox{1.35}{
    \begin{tabular}{|c|c|c|c|}
        \rowcolor[HTML]{b1457c} \textcolor{white}{Algorithm} & \textcolor{white}{type} & \textcolor{white}{$\epsilon$} & \textcolor{white}{$T$} \\
        {Non-smooth} & - & 0.00 & $18.08 \pm 0.85$ \\
        \hline
        {Smooth} & - & 0.00 & $18.15 \pm 0.75$ \\
        \hline \hline
        \multirow{2}{*}{Non-smooth} & \multirow{2}{*}{random} & 0.025& $18.22 \pm 1.28$\\
        & & 0.050& $18.68 \pm 1.97$\\
        \hline
        \multirow{2}{*}{Smooth} & \multirow{2}{*}{random}& 0.025 & $18.17 \pm 0.84$\\
        & & 0.050 & $18.33 \pm 2.23$\\
        \hline \hline
        \multirow{2}{*}{Non-smooth} & \multirow{2}{*}{adv}& 0.025 & $19.33 \pm 1.76$ \\
        & & 0.050& $19.76 \pm 2.85$\\
        \hline
        \multirow{2}{*}{Smooth} & \multirow{2}{*}{adv}& 0.025 & $18.51 \pm 0.65$ \\
        & & 0.050 & $18.84 \pm 1.12$\\
        \hline
    \end{tabular}}
    \caption{Comparison of smooth and non-smooth policies under with no (-), random, or adversarial (Adv) noise at noise level $\epsilon$. $T$ is the time required to complete the task.}
    \label{table:timings}
\end{table}

\section{Conclusion}
In this work, we developed a reinforcement learning approach for optimizing the trajectory of multiple agents for completing a data harvesting mission in two dimensions. We used PPO combined with an adversarially learned regularizer to smooth the policy network to protect against noise and disturbances. From the simulation experiments, we demonstrate that the control policy learned via our proposed PPO method outperforms policies derived from RL using DDQN or planning using the A* algorithm in the multi-agent data harvesting problem. The results also showed that smoothing the policy through regularization improves robustness with respect to disturbances. In future work, we are interested in exploring whether the optimal control analysis can be expanded to find and formally prove an optimal policy (or at least establish a bound on the optimality gap) when working in spaces with dimensions greater than one. Additionally, we intend to investigate other forms of the terminal penalty.

\section*{ACKNOLWEDGEMENTS}
This work was funded in part by NSF through grants ECCS-1931600 and ECCS-1917166.

\bibliographystyle{IEEEtran}
\balance
\bibliography{references.bib}
\end{document}